\documentclass[a4paper,fleqn,10pt]{article}\usepackage[]{graphicx}\usepackage[]{color}
\makeatletter
\def\maxwidth{ %
  \ifdim\Gin@nat@width>\linewidth
    \linewidth
  \else
    \Gin@nat@width
  \fi
}
\makeatother

\definecolor{fgcolor}{rgb}{0.345, 0.345, 0.345}

\usepackage{framed}
\makeatletter
 {\par\unskip\endMakeFramed%
 \at@end@of@kframe}
\makeatother

\definecolor{shadecolor}{rgb}{.97, .97, .97}
\definecolor{messagecolor}{rgb}{0, 0, 0}
\definecolor{warningcolor}{rgb}{1, 0, 1}
\definecolor{errorcolor}{rgb}{1, 0, 0}
\newenvironment{knitrout}{}{} 

\usepackage{alltt}
\usepackage{a4wide}
\definecolor{urlcolor}{rgb}{0,0.5,0}
\definecolor{linkcolor}{rgb}{0.5,0,0}
\definecolor{citecolor}{rgb}{0,0,0.5}
\usepackage[pdfpagemode=UseOutlines,urlcolor=urlcolor,linkcolor=linkcolor,citecolor=citecolor,colorlinks=true]{hyperref}

\usepackage[utf8]{inputenc}
\usepackage[british]{babel}
\usepackage{natbib}
\usepackage{enumerate}

\newcommand{\rinline}[1]{???}

\usepackage{amsmath,amsfonts,amssymb}

\newcommand{\eqn}[1]{\begin{equation}#1\end{equation}}
\newcommand{\term}[1]{\textbf{#1}}

\DeclareMathOperator{\E}{\mathbf{E}}
\DeclareMathOperator{\Var}{\mathbf{Var}}

\newcommand{\cond}{\,\vert\,}

\title{Monitoring the spread of COVID-19 by estimating reproduction numbers over time}
\author{%
    Thomas Hotz$^1$, Matthias Glock$^1$, Stefan Heyder$^1$, Sebastian Semper$^1$,\\ Anne Böhle$^2$, Alexander Krämer$^2$\\[24pt]
    $^1$ Institut für Mathematik, Technische Universität Ilmenau\\
    {\fontsize{10}{12}\selectfont\texttt{\{thomas.hotz,matthias.glock,stefan.heyder,sebastian.semper\}@tu-ilmenau.de}}\\[6pt]
    $^2$ School of Public Health, Bielefeld University\\
    {\fontsize{10}{12}\selectfont\texttt{\{anne.boehle,alexander.kraemer\}@uni-bielefeld.de}}
}
\date{18/04/20}
\IfFileExists{upquote.sty}{\usepackage{upquote}}{}
\begin{document}

\setlength{\parindent}{0pt}
\setlength{\parskip}{6pt}

\maketitle

\begin{abstract}
To control the current outbreak of the Coronavirus Disease 2019, constant monitoring of the epidemic is required since, as of today, no vaccines or antiviral drugs against it are known.
We provide daily updated estimates of the reproduction number over time at \url{https://stochastik-tu-ilmenau.github.io/COVID-19/}.
In this document, we describe the estimator we are using which was developed in \citep{fraser2007}, derive its asymptotic properties, and we give details on its implementation.
Furthermore, we validate the estimator on simulated data, demonstrate that estimates on real data lead to plausible results, and perform a sensitivity analysis.
Finally, we discuss why the estimates obtained need to be interpreted with care.
\end{abstract}

\tableofcontents
\clearpage

\section{Introduction}

As the Coronavirus Disease 2019 (COVID-19) threatens humanity, unprecedented measures to stop its spread have been adopted around the globe.
In many countries, schools have closed and curfews have been imposed.
Given the enormous burden these measures place on the economy, sooner or later they have to be relaxed.
This raises important questions for policymakers and public health specialists.
How large is the effect of these measures?
Do they effectively stop the spread of COVID-19?
What will happen if restrictions get relaxed?
And in the future, how can we see whether the epidemic is getting out of hands again?

To answer these questions, one needs to know how fast the epidemic is growing.
In infectious disease epidemiology, this is measured by the \term{reproduction number}, i.e. the mean number of people someone who got infected will infect in the course of time.
Its \term{critical value} clearly is $1$: for larger values the epidemic will grow, for smaller values it will diminish.

Since conditions may change in the future, e.g. when countermeasures are introduced or lifted, the reproduction number may also change.
We therefore follow \citet{fraser2007} and consider what he calls the \emph{instantaneous} reproduction number $R(t)$ at time $t$, and for which he suggests the estimator
\begin{equation}\label{fraser}
\hat R(t) = \frac{I(t)}{\sum_{\tau=1}^\infty w(\tau) I(t-\tau)}
\end{equation}
where $I(t)$ is the number of incident cases at time $t$ and $w$ specifies the so-called \term{infectivity profile}, i.e. the distribution of the \term{generation time}, which is assumed to be known.
To the best of our knowledge, this estimator has first been published by Fraser and others in \citep{grassly2006}.
An overview of other estimators may be found in \citep{Obadia2012}.

We explain the probabilistic model behind this estimator following \cite[Web Appendix 1]{cori2013} in Section~\ref{DerivEst}.
In addition, we analytically derive asymptotic confidence intervals (with details given in Appendix~\ref{ConfInt}) which are simple to compute.
Here, we differ from \citet{grassly2006} who use computationally more elaborate resampling techniques, namely the bootstrap, to obtain confidence intervals; \citet{cori2013} on the other hand take a Bayesian approach, assuming a certain gamma prior distribution for $R(t)$.

In Section~\ref{SpecCOVID}, some epidemiologically relevant properties of COVID-19 are discussed, and the infectivity profile is modelled.
The estimator and corresponding confidence intervals are validated on simulated data in Section~\ref{ValidSim}.
Then, we apply this methodology to real data for Germany in Section~\ref{ApplReal}, followed by a sensitivity analysis in Section~\ref{SensAnal}.
Finally, the results are summarised in Section~\ref{DiscOut}, also discussing difficulties with this approach.

In order to continuously monitor the spread of COVID-19, a designated website has been created where the results of our analysis are shown and updated daily.
It is available at \url{https://stochastik-tu-ilmenau.github.io/COVID-19/} in English for all affected countries based on the data from \citep{jhu} as well as in German for Germany and its federal states based on the data from \citep{rki} at \url{https://stochastik-tu-ilmenau.github.io/COVID-19/germany}.
The source code for that website as well as for this report may be found at \url{https://github.com/Stochastik-TU-Ilmenau/COVID-19/tree/gh-pages}, rendering this fully reproducible research.
We note that a similar analysis using the Bayesian approach of \citep{cori2013} was presented by \citet{abbott2020} with updates at \url{https://epiforecasts.io/covid/posts/global/}.

\section{Derivation of the estimator}
\label{DerivEst}

The following is an adaptation of the modelling in \citep{fraser2007} and \cite[Web Appendix 1]{cori2013}.

\term{Time} is taken to be discrete, i.e. we consider days $t \in \mathbb Z$, since the spread of the epidemic shows a strong intraday variability (e.g., there are fewer infections during the night when people are at sleep), and the time scales of incubation and infectious period are on the order of days. Also, cases are reported on a daily basis.

The number of \term{incidences}, i.e. newly infected cases, at day $t$ will be given as $I(t)$.
The \term{infection age} of an infected person in days, i.e. the number of day elapsed since the infection, is denoted by $\tau \in \mathbb N_0$.

The spread of the epidemic depends strongly on the time-dependent \term{transmissibility} $\beta(t, \tau) \geq 0$ specifying the expected number of susceptible individuals an infectious person at infection age $\tau$, a so-called \term{primary case}, will infect at time $t$.
The transmissibility is in particular affected by the \term{contact rate}, i.e. the mean number of people an infected person meets per day, and the \term{infectiousness} of the primary case.
The former is addressed by \term{non-pharmaceutical interventions} such as school closures and curfews, the latter is a virological feature of the disease.
Therefore we make a crucial \term{structural assumption}, namely that they separate:
\eqn{ \beta(t, \tau) = R(t)\, w(\tau), \label{struct}}
where $R(t) \geq 0$ denotes the (instantaneous) \term{reproduction number} at time $t$ of transmission, i.e. when the \term{secondary case} gets infected by the primary case, and $w(\tau) \in [0,1]$ specifies the \term{infectivity profile} at infection age $\tau$.
This models the belief that contact rates change over time but the infectiousness of the primary case depends only on $\tau$
which is debatable, however: when rules for isolation or quarantine change are loosened, e.g. because hospital capacities are exhausted, $\beta$ will change differently for different values of $\tau$; we will reiterate this point in Section~\ref{DiscOut}.
It is also reflected in the fact that any constant factor may be alternatively incorporated into $R$ or $w$.
The latter is therefore standardised such that
\eqn{\sum_{\tau = 0}^\infty w(\tau) = 1,}
i.e. $w$ is a \term{probability distribution} which can be interpreted as follows: for a fixed time $t$ randomly pick a pair of individuals where the first one is a primary case that got infected at time $t$, in turn infecting the second one later; $w(\tau)$ is the probability that the second case got infected at time $t+\tau$, i.e. at infection age $\tau$ of the primary case.
$w$ thus specifies the distribution of the \term{generation time}.
It is assumed to be \term{known}; see Section~\ref{SpecCOVID} on how we model it for COVID-19.

In a \term{stochastic model} for the dynamics of the epidemic, $I(t)$ is given as the number of successful transmissions from an infectious person to someone who is susceptible to the disease.
Assuming that each possible transmission succeeds independently (thus ignoring the possibility of multiple infections) with a probability corresponding to $\beta$, and if there are many possible transmissions, $I(t)$ is -- by the law of small numbers -- approximately \term{Poisson distributed} conditional on the past. The intensity of this Poisson distribution is equal to
\eqn{ \E(I(t) \cond I(t-1), \dots) = \sum_{\tau=1}^\infty \beta(t,\tau) I(t-\tau) = R(t) \sum_{\tau=1}^\infty w(\tau) I(t-\tau)\,. \label{condE}}
Here, transmissions \term{on the same day} are ruled out, i.e. $w(0) = 0$, which is a realistic assumption since the incubation period will be at least one day.

The last equation suggests the \term{estimator} $\hat R(t)$ for $R(t)$ given in \cite[Equation (9)]{fraser2007},
\eqn{ \hat R(t) = \frac{I(t)}{\sum_{\tau=1}^\infty w(\tau) I(t-\tau)}\,. \label{estim}}

Note that the \term{case reproduction number} $R_c(t)$, i.e. the expected number of people a primary case infected at time $t$ will infect, is given by, cf. \cite[Equations (2) and (8)]{fraser2007},
\eqn{ R_c(t) = \sum_{\tau=1}^\infty \beta(t+\tau, \tau) = \sum_{\tau=1}^\infty R(t+\tau)\,w(\tau)\,. }
This is of course difficult or even impossible to estimate as it depends on future contact rates, i.e. on countermeasures that will be imposed.
However, assuming that conditions remain the same in the future, i.e. $R(s) = R(t)$ for $s > t$, we obtain $R(t)$ again, cf. \cite[Equation (3)]{fraser2007},
\eqn{ \sum_{\tau=1}^\infty R(t+\tau) w(\tau) = R(t) \sum_{\tau=1}^\infty w(\tau) = R(t)\,.}
This explains why $R(t)$ is called \emph{reproduction number}.

For large intensities, i.e. if the conditional expectation in Equation~\eqref{condE} is large, the distribution of $\hat R(t)$ can be well approximated by a Gaussian distribution, with small standard errors.
From this, \term{asymptotic confidence intervals} can be derived, see Section~\ref{ConfInt}.
If $q$ denotes the $(1-\frac\alpha 2)$-quantile of the standard normal distribution then
\eqn{ \Biggl[\hat R(t) - q \sqrt{\frac{\hat R(t)}{\sum_{\tau=1}^\infty w(\tau) I(t-\tau)}}, \hat R(t) + q \sqrt{\frac{\hat R(t)}{\sum_{\tau=1}^\infty w(\tau) I(t-\tau)}} \Biggr] }
is an (asymptotic) $(1-\alpha)$-confidence interval for $R(t)$.
Note that in practice ten or more incident cases should suffice for the asymptotics to be reliable.

\section{Specifics of COVID-19}
\label{SpecCOVID}

As COVID-19 is a new disease, first being described at the end of 2019, its virological features have not yet been conclusively determined.
Nonetheless, we tried to choose parameters in agreement with the current state of research.

For comparisons, we note that in a population without any countermeasures, the \term{basic reproduction number} $R_0$ is believed to be given by some value between 2.4 and 4.1 \citep{read2020}.

The \term{incubation time}, i.e. the time from infection until symptom onset, ranges from 1 to 14 days with a mean of 5 to 6 days; the virus can be detected from 1 to 2 days before symptom onset for up to 7 to 12 days in moderate cases, and even up to two weeks in severe cases \citep{who2020}.
We therefore may indeed assume that $w(0) = 0$.

For modelling the infectivity profile $w$, it is important to realise that it is not proportional to the amount of viral specimens that can be detected in an infected person's sputum, say.
Indeed, since severe cases are very likely to be hospitalised and thus strictly isolated, the probability of infecting someone more than 12 days after infection is very low.
Similarly, before symptom onset the probability for transmission might be very low since no sputum is spread.

\begin{figure}[!p]
\begin{knitrout}
\definecolor{shadecolor}{rgb}{0.969, 0.969, 0.969}\color{fgcolor}
\includegraphics[width=\maxwidth]{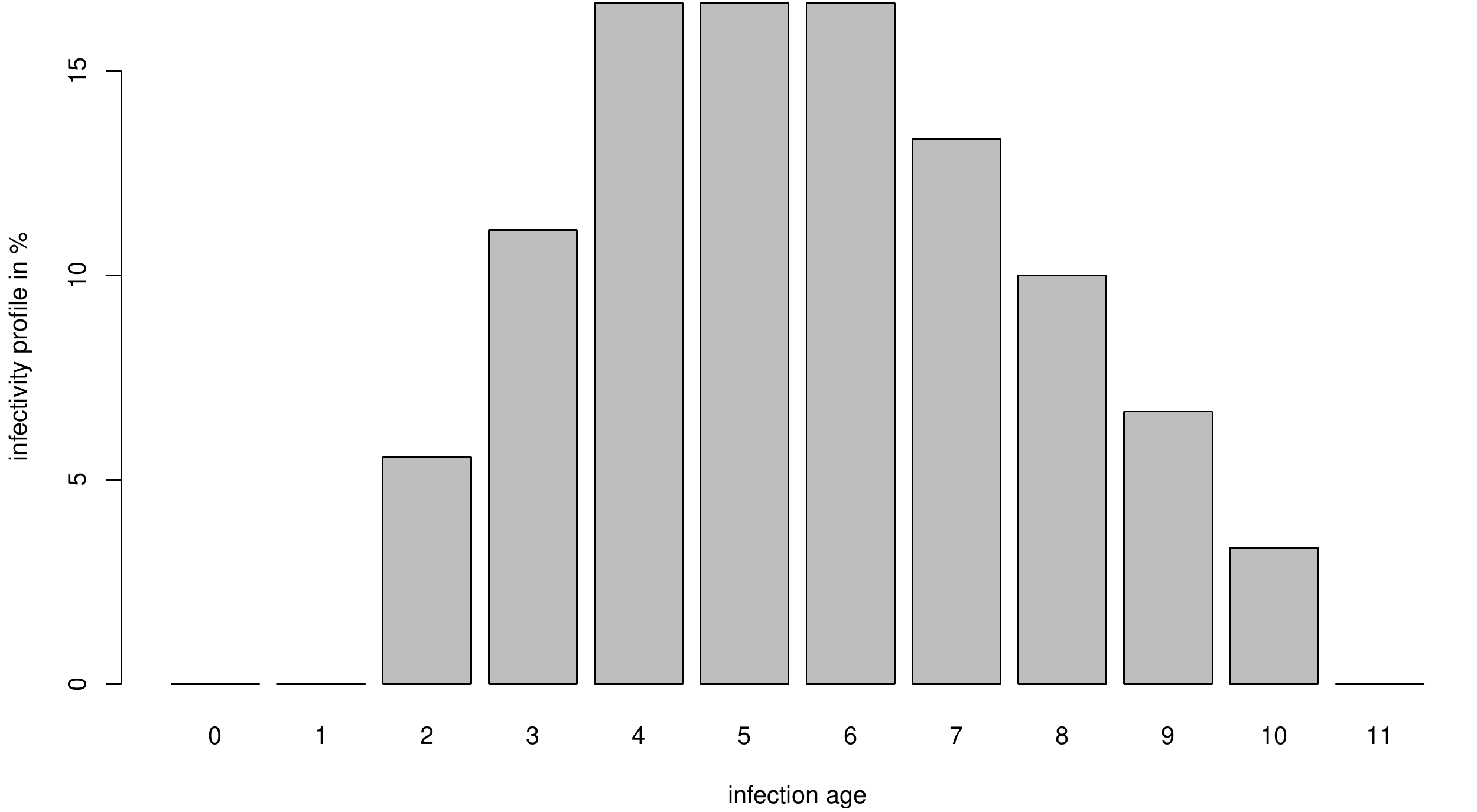} 

\end{knitrout}
\caption{Modelled infectivity profile $w$.}\label{infProf}
\end{figure}

The infectivity profile is therefore set to start with $0$ on the first day after infection with a linear increase up to day 4, remaining constant up to day 6 and decaying linearly again until being $0$ at day 11; see Figure~\ref{infProf}.
In Section~\ref{SensAnal} we discuss the effect this choice has on the analysis.

\section{Validation on simulated data}
\label{ValidSim}

\begin{figure}[!p]
\begin{knitrout}
\definecolor{shadecolor}{rgb}{0.969, 0.969, 0.969}\color{fgcolor}
\includegraphics[width=\maxwidth]{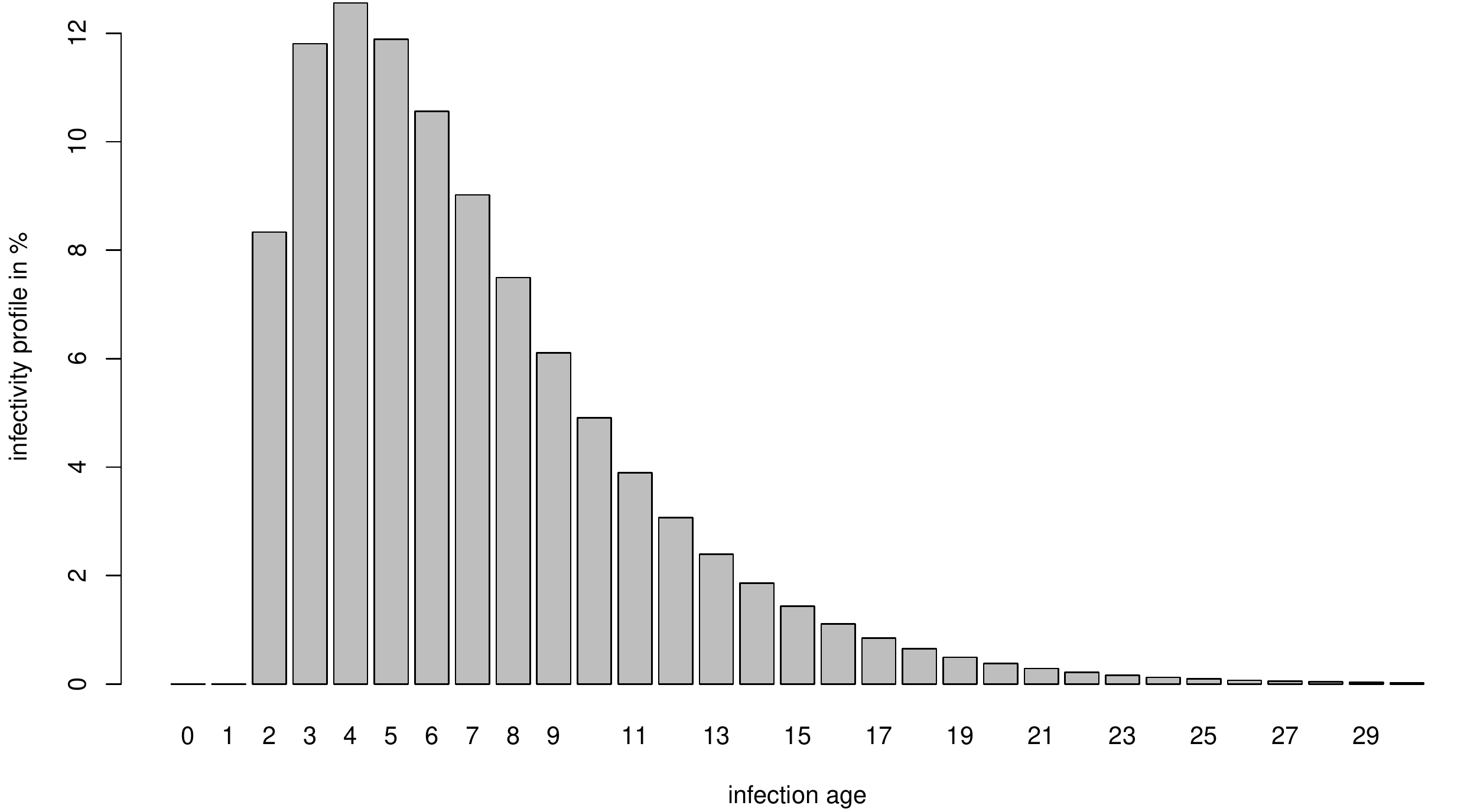} 

\end{knitrout}
\caption{Computed infectivity profile $w$ corresponding to the simulation.}\label{simPlotInfect}
\end{figure}

To validate the estimator, we simulate a \term{stochastic SEIR} (a.k.a. Kermack-McKendrick) \term{model}.
To be more precise, we consider a discrete-time Markov chain describing a population of $n = 1$~million people with each individual being in one of four states: \term{susceptible}, i.e. not yet infected; \term{exposed}, more aptly called \term{latent}, i.e. infected but not yet infectious; \term{infectious}; or \term{recovered} and thus immune.
We start at time $0$ with $100$ latent individuals, all others initially being susceptible.
At each time step, a susceptible person becomes infected if the virus is transmitted through contact with an infectious person; such contacts happen independently with probability $p_E$. A latent person becomes infectious with probability $p_I$, and an infectious person recovers with probability $p_R$; otherwise an individual remains in its state.

This results in incubation times, i.e. times spent in the latent state, which are geometrically distributed with mean $1/ {p_I}$; for this to be $3$, we set $p_I = 1/ {3}$.
Similarly, the infectious period is geometrically distributed with mean $1/ {p_R}$ which we would like to be $4$, so we set $p_R = 1/ {4}$.
The corresponding infectivity profile $w$ is then given by the convolution of these two geometric convolutions.
It can be calculated analytically, see Appendix~\ref{InfProfSEIR} for details; the result is shown in Figure~\ref{simPlotInfect}.
Note that $w(1) = w(0) = 0$ since it takes at least one day to become latent and another one to become infectious in this model.

The basic reproduction rate is then given by $R_0 = {n p_E}/{p_R}$ since an infected person on average infects $n p_E$ individuals per day (if all were susceptible) for $1/ {p_R}$ days on average.
In order to simulate an epidemic with $R_0 = 2.5$, we set $p_E = {R_0 p_R}/n$ accordingly.

Over time, the reproduction number changes naturally because more people recover and become immune: $R(t)$ is $R_0$ times the proportion of susceptible individuals at that time.
In addition, we assume that countermeasures have been imposed at time $30$, resulting in $R(t)$ being $0.7$ times the proportion of susceptibles afterwards, and that measures have been relaxed at time $50$, resulting in $R(t)$ being $1.3$ times the proportion of susceptibles thereafter.

\begin{figure}[!p]
\begin{knitrout}
\definecolor{shadecolor}{rgb}{0.969, 0.969, 0.969}\color{fgcolor}
\includegraphics[width=\maxwidth]{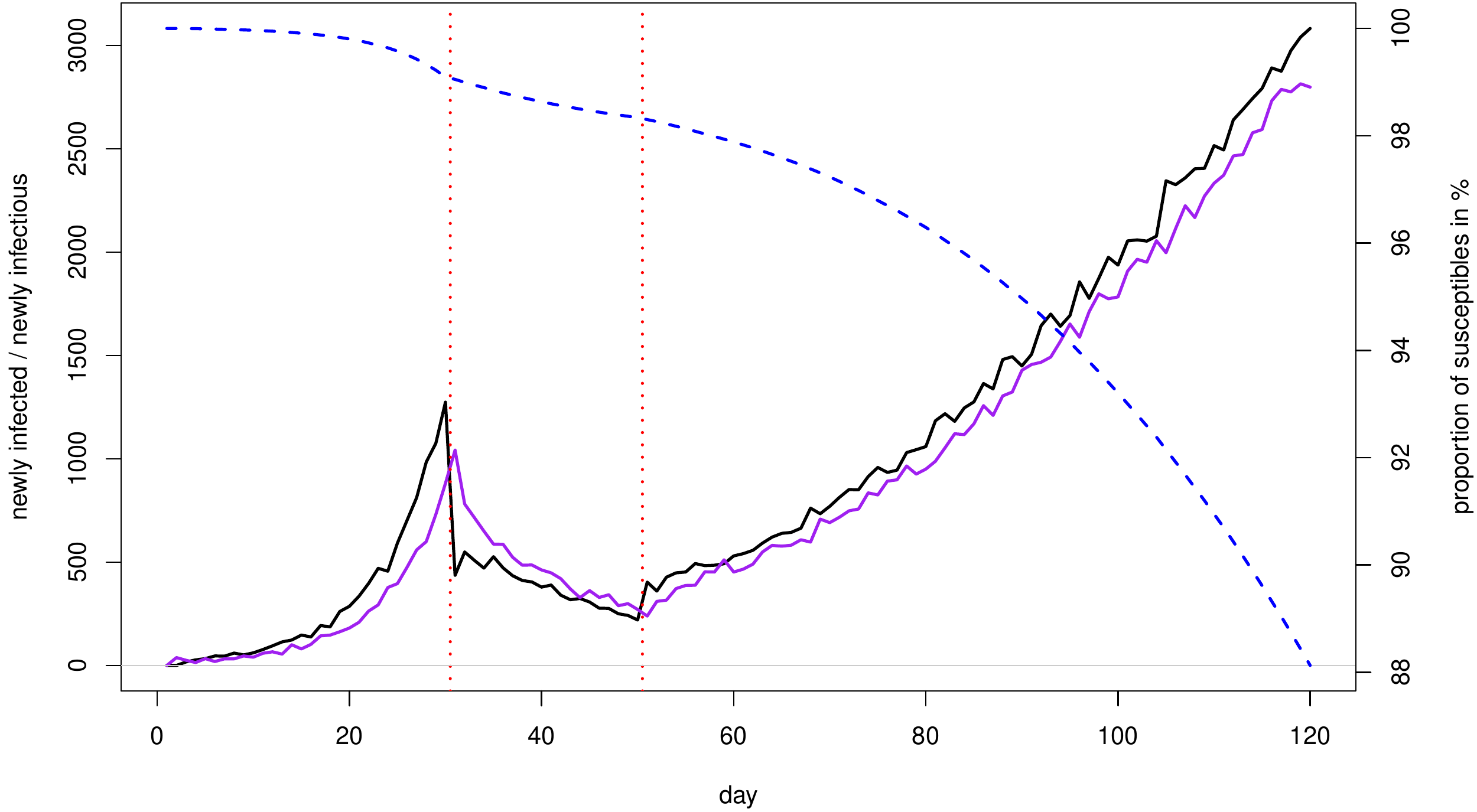} 

\end{knitrout}
\caption{One simulation of the SEIR model; black solid line (left axis): newly infected; purple solid line (left axis): newly infectious; dashed blue line (right axis): proportion of susceptibles; vertical red dotted lines: intervention times.}\label{simPlotSEIR}
\end{figure}

\begin{figure}[!p]
\begin{knitrout}
\definecolor{shadecolor}{rgb}{0.969, 0.969, 0.969}\color{fgcolor}
\includegraphics[width=\maxwidth]{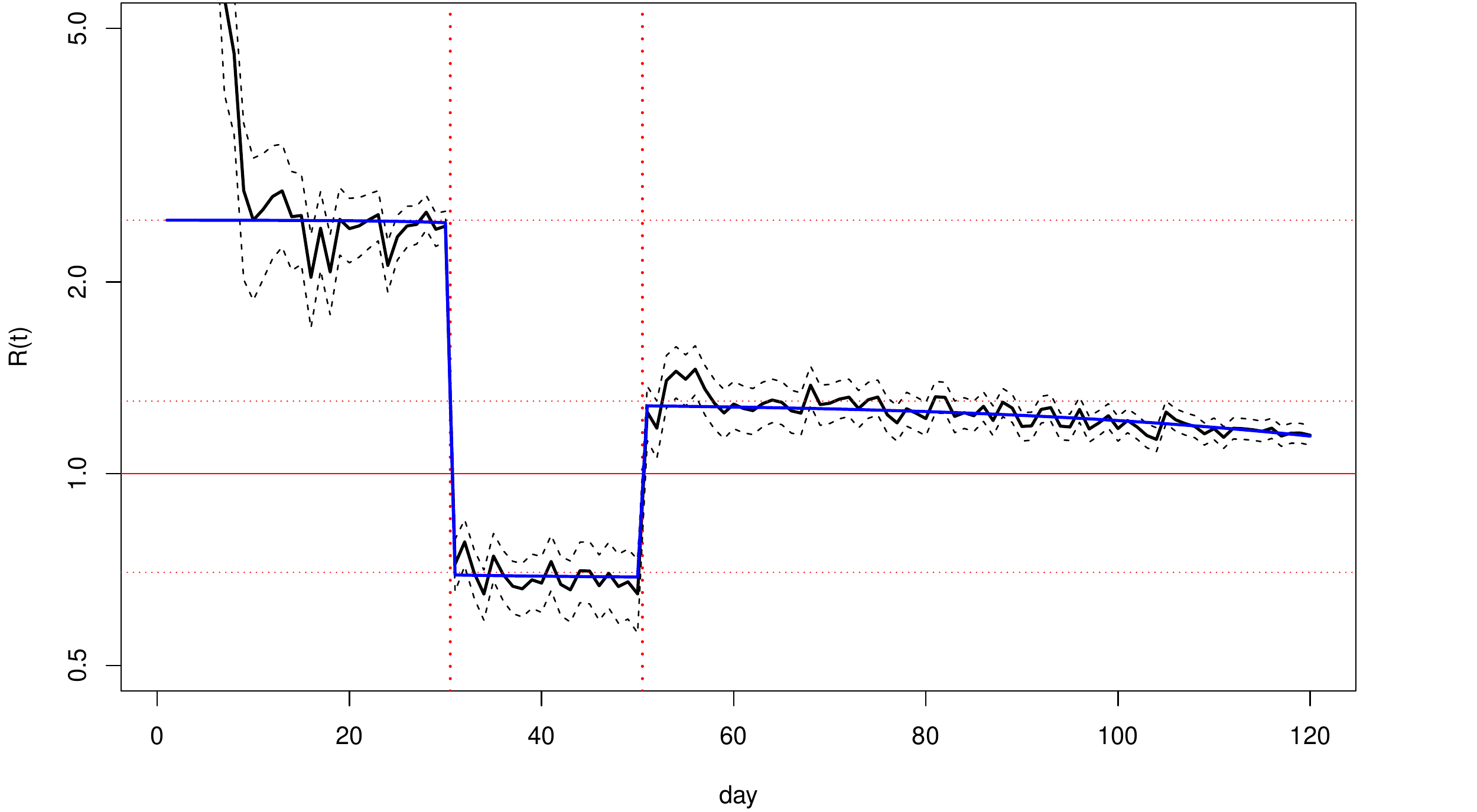} 

\end{knitrout}
\caption{One simulation of the SEIR model; black solid line: $\hat R(t)$; black dashed lines: pointwise 95\%-confidence intervals; blue solid line: $R(t)$; vertical red dotted lines: intervention times; horizontal red dotted lines: corresponding reproduction numbers without decrease in susceptibles taken into account.}\label{simPlotSingle}
\end{figure}

Figure~\ref{simPlotSEIR} shows one simulation run.
The resulting estimates $\hat R(t)$ and pointwise 95\%-confidence intervals ($\alpha = 5\%$ as usual) can be compared with the true reproduction rate $R(t)$ in Figure~\ref{simPlotSingle}.

\begin{figure}[!p]
\begin{knitrout}
\definecolor{shadecolor}{rgb}{0.969, 0.969, 0.969}\color{fgcolor}
\includegraphics[width=\maxwidth]{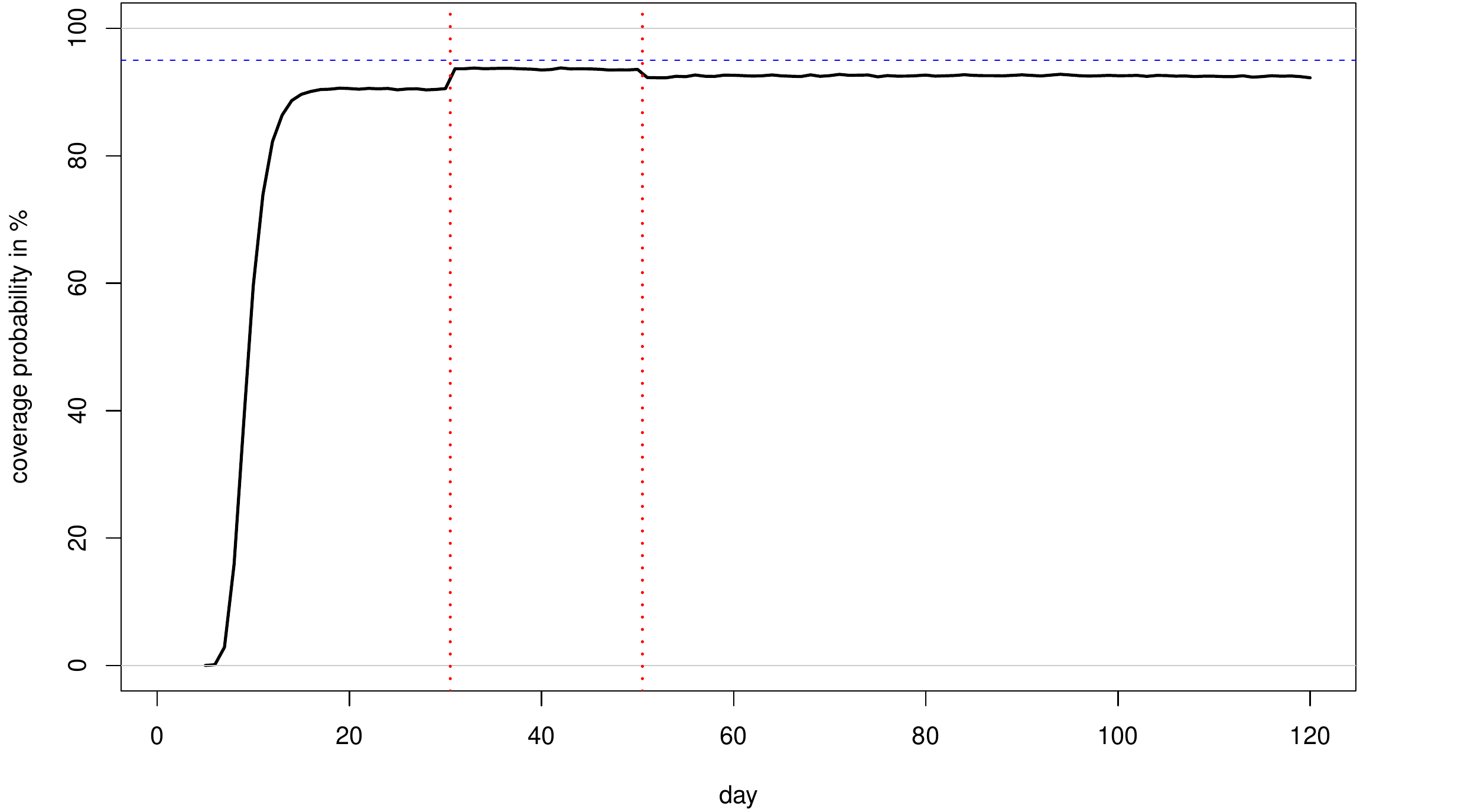} 

\end{knitrout}
\caption{Estimated coverage probability based on $10^{5}$ simulations (black solid line); horizontal blue dahed line: nominal coverage (95\%); vertical red dotted lines: intervention times.}\label{simCover}
\end{figure}

The simulation has been repeated $10^{5}$ times, and for each time point the proportion of confidence intervals containing the true reproduction number has been determined, see Figure~\ref{simCover}.
They appear not quite to have the desired nominal coverage but given that they are only asymptotic confidence intervals, and modelling errors are typically much larger, we consider them acceptable in practice.

These simulations demonstrate how well the estimator is able to detect changes in the reproduction number.
From a practical viewpoint, this is an overly optimistic result. 
In fact, Equation~\eqref{condE} and consequently the estimator $\hat R(t)$ in Equation~\eqref{estim} are based on the number of newly infected cases.
But \term{infection dates} are rarely known.
Instead, cases are reported when they are tested with a positive test result.
In our simple simulation, one should therefore consider the \term{newly infectious cases} at day $t$ as input data $I(t)$ for the estimator.
Note that their increase lags behind the one of the \term{newly infected cases}, i.e. the \term{newly latent cases}, by the incubation time, see Figure~\ref{simPlotSEIR} where they lag behind by about $1$ day, the mode of the incubation time distribution.

\begin{figure}[!p]
\begin{knitrout}
\definecolor{shadecolor}{rgb}{0.969, 0.969, 0.969}\color{fgcolor}
\includegraphics[width=\maxwidth]{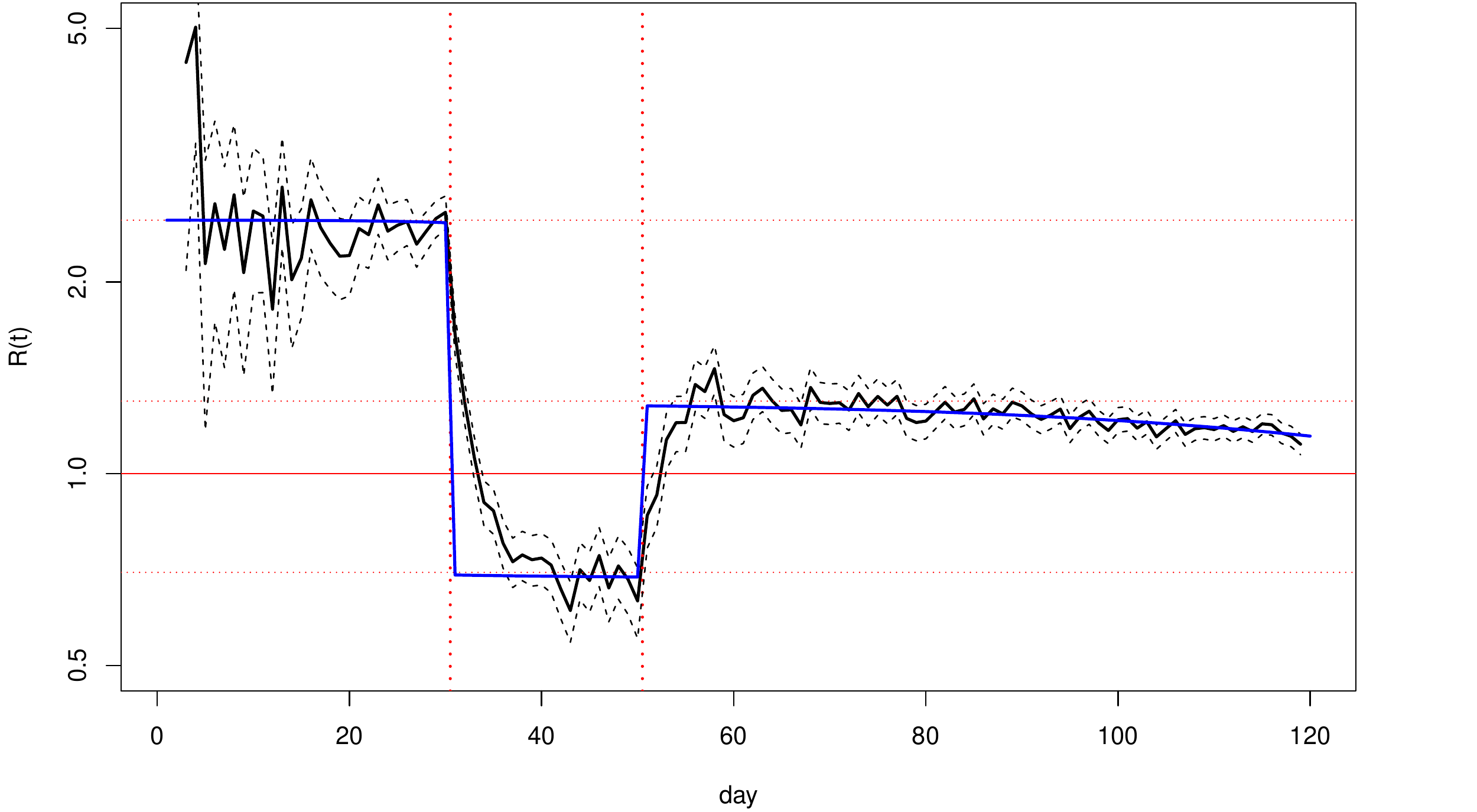} 

\end{knitrout}
\caption{Estimator based on the newly infectious of one simulation of the SEIR model shifted by 1 day; black solid line: $\hat R(t)$; black dashed lines: pointwise 95\%-confidence intervals; blue solid line: $R(t)$; vertical red dotted lines: intervention times; horizontal red lines: corresponding reproduction numbers without decrease in susceptibles taken into account. This is to be compared with Figure~\ref{simPlotSingle} where the estimator is based on the newly infected cases.}\label{simPlotSingleInfectious}
\end{figure}

We use a na\"ive approach to deal with this which we call \term{infection-to-observation period}: we shift the estimator back by the observed lag, i.e. by 1~day.
The result is shown in Figure~\ref{simPlotSingleInfectious} where the jump in $R(t)$ leads only to a rapid change of $\hat R(t)$, approaching the true value $R(t)$ exponentially fast, though.
For real data, the infection-to-observation period is even larger, since symptomatic cases are usually not reported immediately. This will be taken into account in the following section.

\section{Application to real data}
\label{ApplReal}

\begin{figure}[!p]
\begin{knitrout}
\definecolor{shadecolor}{rgb}{0.969, 0.969, 0.969}\color{fgcolor}
\includegraphics[width=\maxwidth]{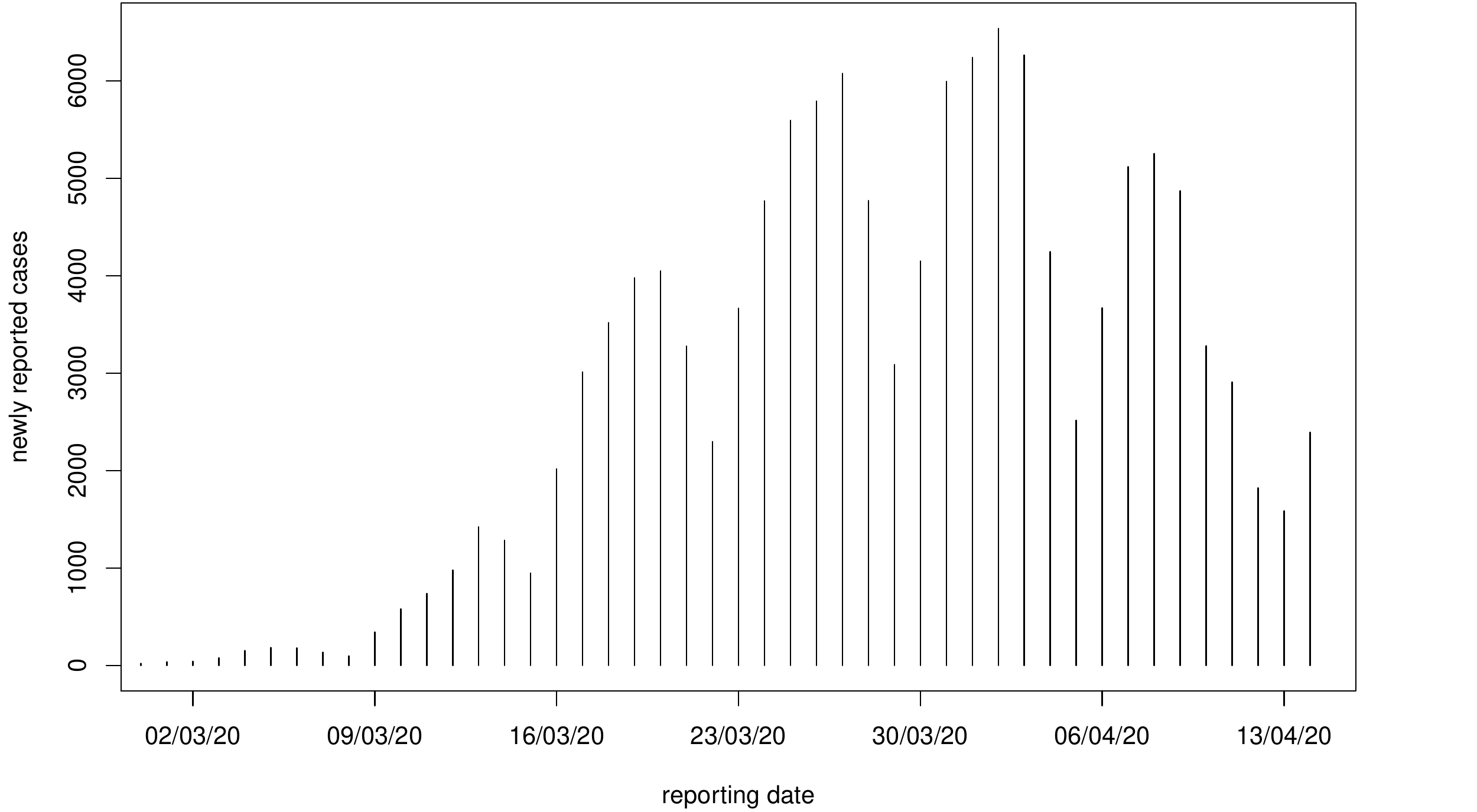} 

\end{knitrout}
\caption{Newly reported cases for Germany over time, based on data from the \citet{rki}.}\label{bund}
\end{figure}

\begin{figure}[!p]
\begin{knitrout}
\definecolor{shadecolor}{rgb}{0.969, 0.969, 0.969}\color{fgcolor}
\includegraphics[width=\maxwidth]{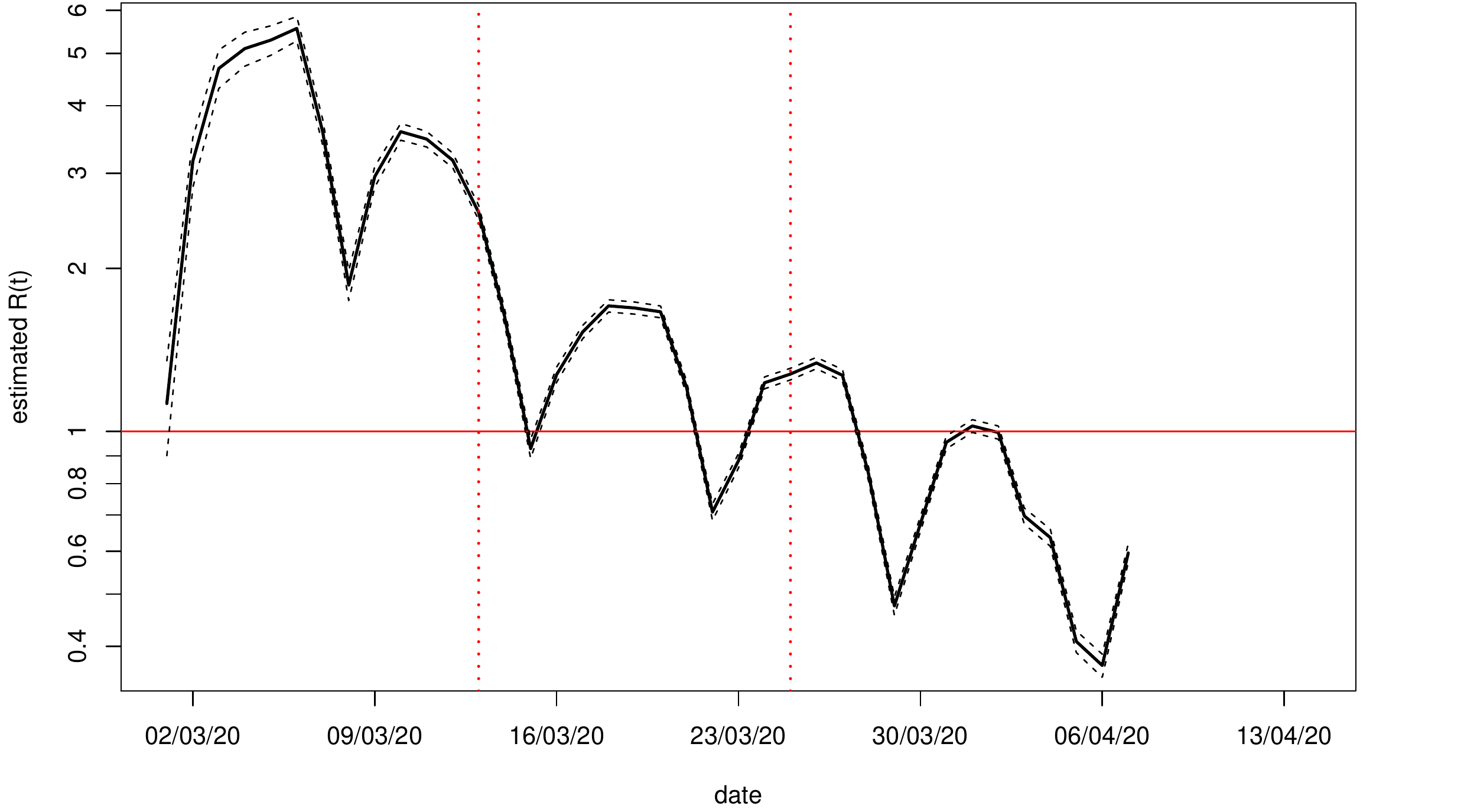} 

\end{knitrout}
\caption{Estimated reproduction numbers for Germany over time (solid line) with pointwise 95\%-confidence intervals (dashed lines); vertical red dashed lines indicate the time period over which countermeasures have been implemented, cf. Table~\ref{npi}.}\label{bundEst}
\end{figure}

As an example, we consider data for Germany and its federal states (Bundesländer) provided by the \citet{rki}, see Figure~\ref{bund} for the total daily reported cases.
Each case in this dataset is labelled with a \term{reporting date}, i.e. the day when the local health authority (Gesundheitsamt) has been notified about the case.
Of course, this is not the day of symptom onset, let alone the day of infection which is needed for the estimator in Equation~\eqref{estim}.
We therefore set an \term{infection-to-observation period} by which we backdate the cases.
It is pragmatically chosen as 5~days of incubation time (cf. Section~\ref{SpecCOVID}) plus 2~more days \term{reporting delay} for testing etc., i.e. the infection-to-observation period is set to 7~days.

Since cases are reported to local health authorities, then collected at the level of states who in turn report them to the federal Robert Koch-Institut, they appear in the dataset a few days later, although with the date of reporting to the local health authority.
Therefore, we exclude data from yesterday and the two days before.

\begin{table}[!t]
\begin{center}
\begin{tabular}{l|l}
\textbf{date of implementation} & \textbf{measure} \\[3pt]
13--18/03/2020 (mostly 16/03/2020) & school closures \\
14--22/03/2020 (mostly 16--22/03/2020) & closure of institutions, restaurants etc. \\
20--25/03/2020 (mostly 22/03/2020) & contact restrictions
\end{tabular}
\end{center}
\caption{Summary of starting dates for non-pharmaceutical interventions introduced by federal states in Germany.}\label{npi}
\end{table}

Based on the backdated data and the infectivity profile from Section~\ref{SpecCOVID} (see Figure~\ref{infProf}), we estimated the reproduction numbers for Germany over time, see Figure~\ref{bundEst}.
Note that there are no estimates for the last 7~days for which new cases are shown in Figure~\ref{bund} due to the infection-to-observation period.

Starting with Bremen on 13/03/2020, more and more restrictive non-pharmaceutical countermeasures have been adopted by the federal states; see Table~\ref{npi} for a short overview.
Their effect on the reproduction number is clearly visible in Figure~\ref{bund}, resulting in a reproduction number of less than 1 with all measures in place.

The strong weekly pattern in the estimates is due to the fact that less cases are reported around weekends, cf. Figure~\ref{bund} where Mondays are marked on the horizontal axis.
We do not compute an average over a sliding window of seven days so the viewer immediately recognizes the size of such artefacts, warning her to be overly confident in the results.
In fact, these artefacts are much larger than the statistical uncertainty due to the stochastic nature of the epidemic which is reflected in the confidence intervals.

\section{Sensitivity analysis}
\label{SensAnal}

\begin{figure}[!p]
\begin{knitrout}
\definecolor{shadecolor}{rgb}{0.969, 0.969, 0.969}\color{fgcolor}
\includegraphics[width=\maxwidth]{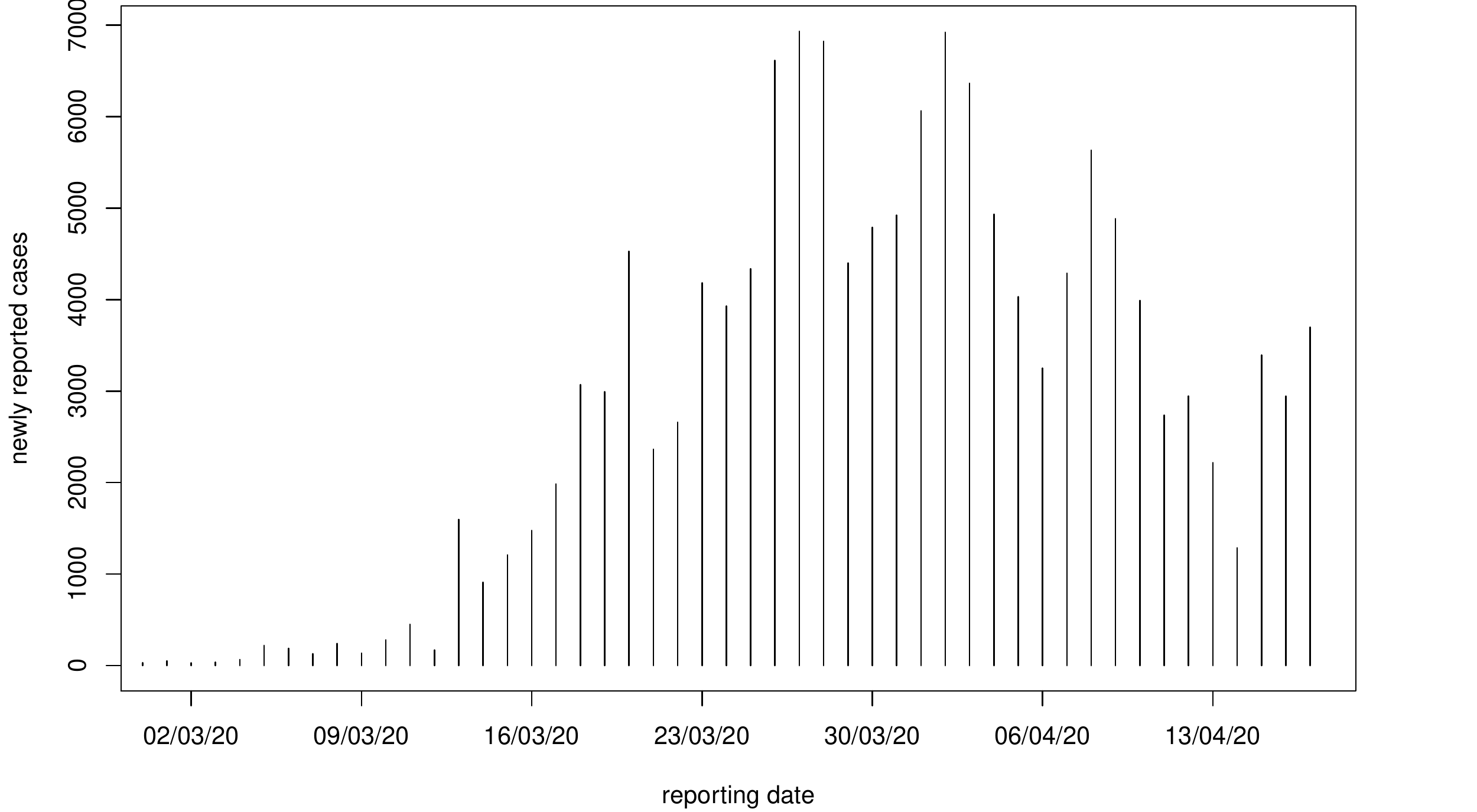} 

\end{knitrout}
\caption{Newly reported cases for Germany over time, based on data from the \citet{jhu}; compare with Figure~\ref{bund}.}\label{germany}
\end{figure}

The estimator depends on two ingredients, the data, of course, and the infectivity profile.
For Germany, there is a second dataset provided by \citet{jhu} whose source are mainly official data, too, but collected at the local level once they are available.
In particular, as soon as data become available, they are marked as reported on that very day.
They therefore show a far less pronounced weekday effect than the data from the \citet{rki}, see Figure~\ref{germany} and compare with Figure~\ref{bund}.
Moreover, data are not backedited, so even yesterday's data are final and can be used.

\begin{figure}[!p]
\begin{knitrout}
\definecolor{shadecolor}{rgb}{0.969, 0.969, 0.969}\color{fgcolor}
\includegraphics[width=\maxwidth]{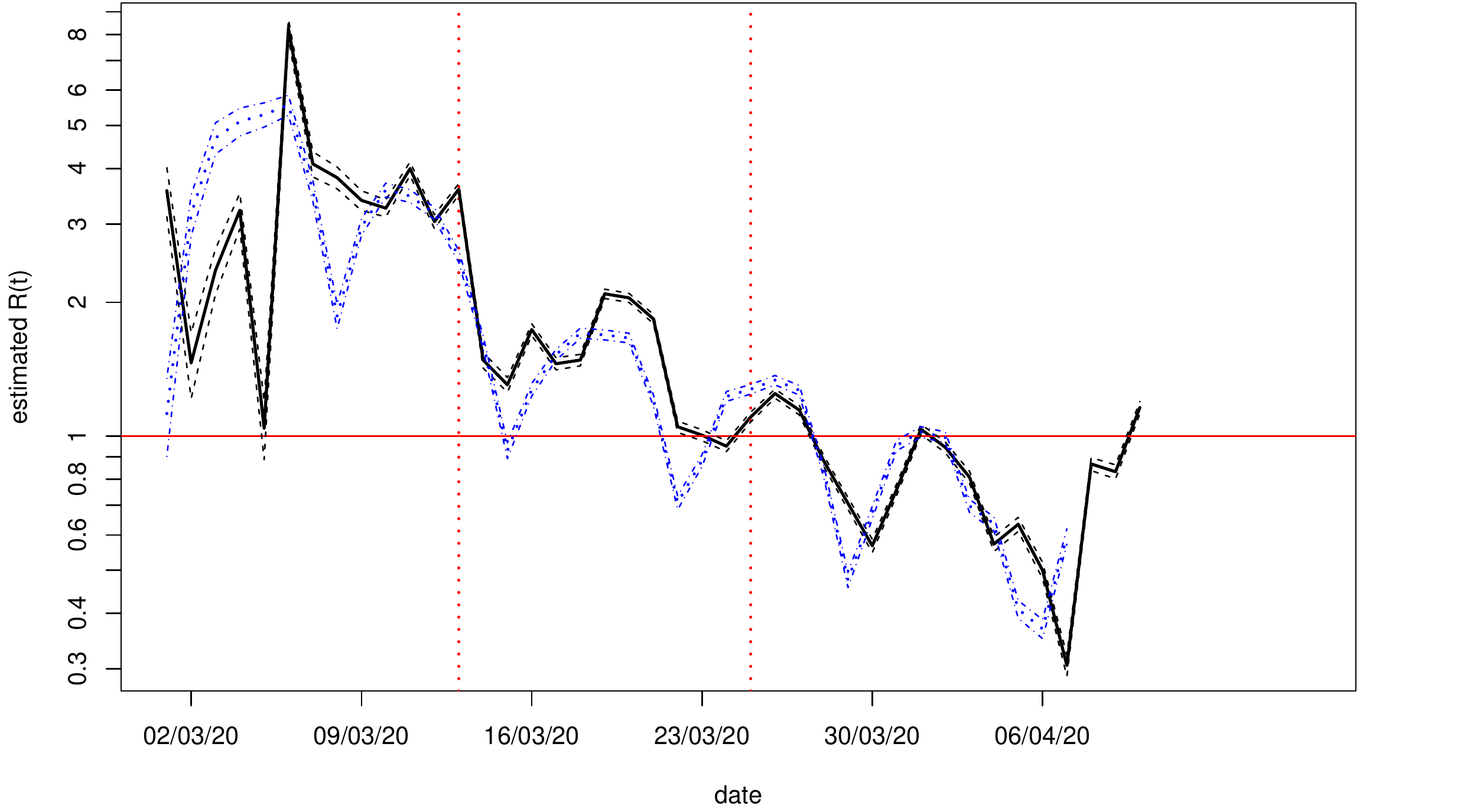} 

\end{knitrout}
\caption{Estimated reproduction numbers for Germany over time (solid/dotted lines) with pointwise 95\%-confidence intervals (dashed/dash-dotted lines) based on data from \citet{jhu}, shown in black (solid), and \citet{rki}, shown in blue (dotted), respectively; vertical red dashed lines indicate the time period over which countermeasures have been implemented, cf. Table~\ref{npi}.}\label{germanyEst}
\end{figure}

The estimates based on the data from the \citet{jhu} differ quantitatively but not qualitatively from the ones using the data of the \citep{rki}, see Figure~\ref{germanyEst}.

\begin{figure}[!p]
\begin{knitrout}
\definecolor{shadecolor}{rgb}{0.969, 0.969, 0.969}\color{fgcolor}
\includegraphics[width=\maxwidth]{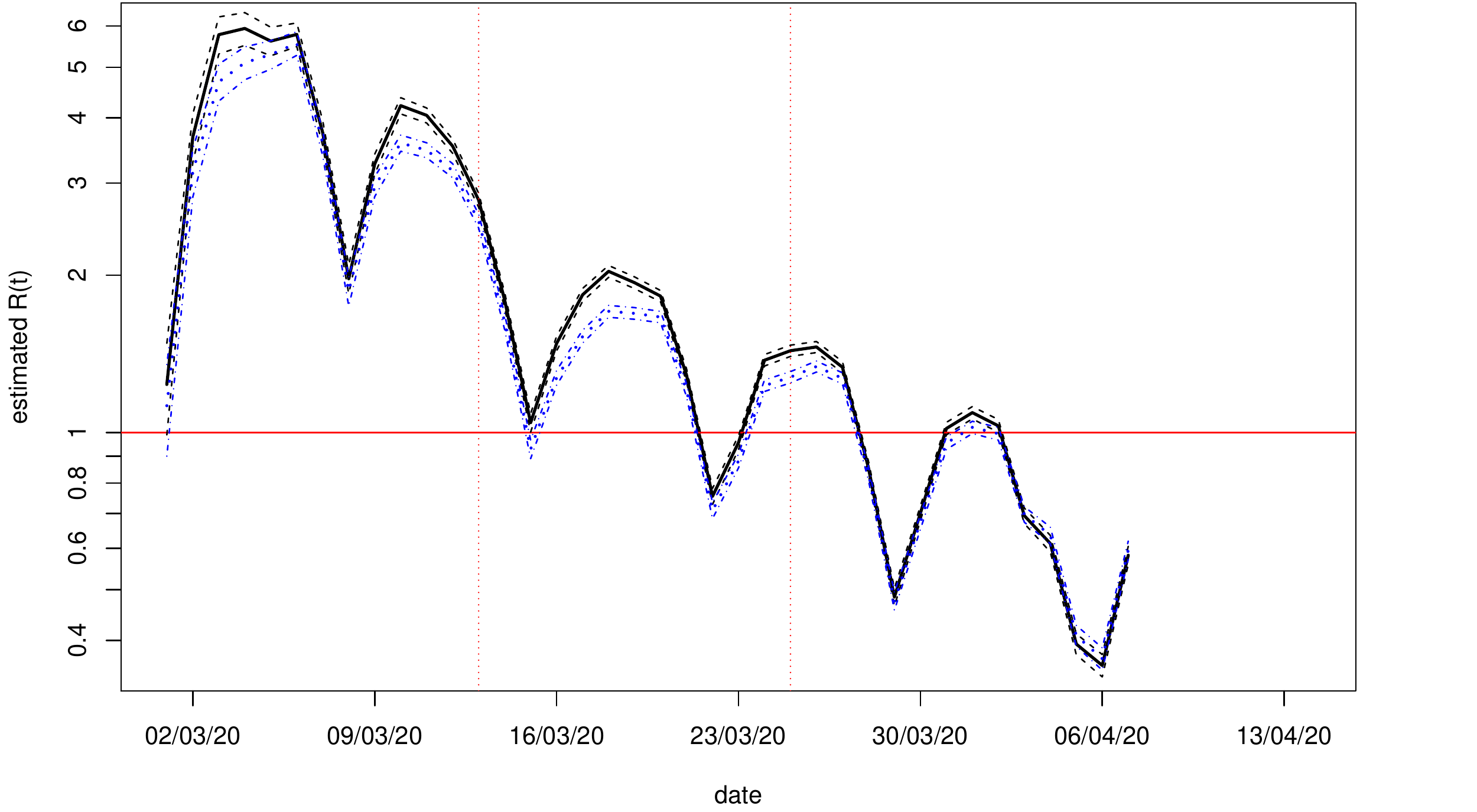} 

\end{knitrout}
\caption{Estimated reproduction numbers for Germany over time (solid/dotted lines) with pointwise 95\%-confidence intervals (dashed/dash-dotted lines) based on data from \citet{rki} using the infectivity profile of the SEIR-model in Section~\ref{ValidSim} (see Figure~\ref{simPlotInfect}), shown in black (solid), and using the infectivity profile modelled in Section~\ref{SpecCOVID} (see Figure~\ref{infProf}), shown in blue (dotted), respectively; vertical red dashed lines indicate the time period over which countermeasures have been implemented, cf. Table~\ref{npi}.}\label{bundSens}
\end{figure}

To understand the effect the infectivity profile exerts on the estimates, we consider the infectivity profile we computed for the stochastic SEIR-model used for simulations in Section~\ref{ValidSim} (see Figure~\ref{simPlotInfect}), employing it to estimate the reproduction numbers using the data from the \citet{rki} again.
When comparing the results with the ones obtained using the infectivity profile modelled in Section~\ref{SpecCOVID} (see Figure~\ref{infProf}), one observes that the former profile has a longer tail, so the estimator takes values from further in the past into account, which for increasing case numbers reduces the denominator in Equation~\eqref{estim}, and hence somewhat increases the estimates.
Once case numbers stabilise, this effect obviously vanishes.
As with the data source, the influence of the infectivity profile appears to be small enough not to matter qualitatively.

\section{Discussion}
\label{DiscOut}

The results for simulated data in Section~\ref{ValidSim} demonstrate the validity of the estimator, and of the asymptotic confidence intervals we derived.
This is substantiated further by the fact that the estimated reproduction numbers' decrease for Germany correlate strongly with enforcement of non-pharmaceutical countermeasures there.

Let us stress the advantage of this estimator over approaches which determine growth rates or doubling times by fitting \term{exponential growth models} to numbers of either new cases or total cases in the initial phase of the epidemic where the proportion of susceptibles is close to $100\%$, cf. \citep{Obadia2012}.
In fact, the latter models are implicitly based on the assumption that conditions do not change such that the epidemic spreads with a constant growth rate, and thus with a constant reproduction number.
But here we aim to determine a varying reproduction number.
Fitting exponential growth models to total case numbers therefore is not conducive, even when localising the procedure by considering short time windows.
Indeed, the case numbers from the past which occurred under different conditions will always affect the estimates.
This problem is alleviated when exponential growth models are fitted locally to the numbers of new cases.
Still, one needs to assume that conditions change slowly -- which is debatable for the drastic measures which have been implemented quickly.
In any case, even if one could observe new infections directly, the resulting estimates would be (additionally) smoothened, as opposed to the unbiased, sharp results obtained for the estimator we consider here (cf. Figure~\ref{simPlotSingle}).

Nonetheless, the estimates have to be cautiously interpreted.
For one, the calculated confidence intervals quantify only a rather small part of the uncertainty, namely the one which stems from the stochastic nature of the epidemic's evolution over time.
Other uncertainties may affect the estimates much more, in particular when case numbers are large.
In the following, we discuss those which we believe to be most important.

The first set of difficulties concerns the \term{quality of the data}.
\begin{enumerate}[(a)]
\item Not all infections are reported, for example because cases remain asymptomatic, or because infected persons die without having been tested.
If the proportion of infections which get reported stays constant over time (or at least varies slowly), both numerator and denominator in Equation~\eqref{estim} are multiplied by the same factor, so they cancel and the estimates are not affected.
Changes in the reporting or testing methodology, however, will affect the estimates, as they will be indistinguishable from a true increase or decrease in the number of infections.
This will for example happen if more people are tested due to higher capacities in testing facilities, or if lower case numbers allow more extensive tests of potential contacts, or if deaths are attributed to the disease without testing when such capacities are exhausted.
Potential remedies include to not only consider reported infections but take fatalities, test rates etc. into account.
\item The reporting date is not the date of infection: when the patient becomes symptomatic, he has to visit a physician, samples have to be tested, the test results need to be interpreted, and finally reported to the authorities.
The strength of this effect is visible from the periodic pattern related to the days of the week in Figures~\ref{bund} and~\ref{bundEst}.
For some of the data provided by the \citet{rki}, both the reporting date, and the day of symptom onset are known, which in principle allows to infer dates of symptom onset for the entire dataset, thus getting rid of the weekday's influence.
But the difficulty that the estimator is based on knowing the date of infection remains, cf. Section~\ref{ValidSim} with Figure~\ref{simPlotSingleInfectious}.
To treat this properly, one would need to know the distribution of the incubation times, and compute a deconvolution.
\item Imported cases, i.e. travellers who became infected abroad and got reported after returning home, should not be counted as secondary cases because the corresponding primary case has not been accounted for. However, the location of infection is often unknown; such cases will then unduly increase the numerator in Equation~\eqref{estim}, and hence also the estimated reproduction number.
This might explain the surprisingly large estimated values -- larger than 4, cf. Section~\ref{SpecCOVID} --  at the beginning of March in Germany, see Figure~\ref{bundEst}, when many infections were acquired during holidays abroad.
\end{enumerate}

Other problems originate from the modelling approach.
\begin{enumerate}[(a)]\setcounter{enumi}{3}
\item In Equation~\eqref{struct}, a structural assumption was made: the infectivity profile does not change over time.
If changing conditions affect cases at different infection ages differently, e.g. because the health system is overwhelmed and no longer can provide for high quality isolation of severe cases (with higher infection ages), or because better medical treatment for such cases becomes available, then the change of the transmissibility $\beta$ depends on the infection age.
As a result, the estimates for the reproduction number will combine the changes for the different infection ages into a certain average.
\item Similarly, the method does not distinguish individuals in different strata of the population, e.g. age groups or regions.
So changes which affect certain strata more and others less, e.g. school closures, will again be averaged over the population.
\item Finally, the infectivity profile requires modelling.
We stress that this needs to be distinguished from a virological assessment of a case's level of infectiousness, as it rather describes the potential to successfully transmit the virus.
For example, the probability at a late stage of the infection may be assumed to be very low: such a person is most likely to be well isolated, either at home (where either all other members of the household have already been infected or apparently are immune) or at a hospital (where isolation measures are strict), so even though from a virological point of view the person may be highly infectious, she probably will not cause a secondary infection at that stage.
From data on chains of infection, it may be possible to directly estimate the infectivity profile from the observed generation times.
\end{enumerate}

All these issues render comparisons between countries particularly difficult.
In any case, they sound a note of caution when looking at the proposed estimates.
Nonetheless, encouraged by the results in Section~\ref{ApplReal}, we believe that qualitatively correct conclusions may be drawn from them, and that they may prove useful to continuously monitor the spread of COVID-19.

\paragraph{Acknowledgements.}\addcontentsline{toc}{section}{Acknowledgements}

The authors thank Dr. med. Luise Prüfer-Krämer, Steering Committee Member of the German Society of Tropical Medicine and Global Health and practising physician, for many fruitful discussions and insights into the care of COVID-19 patients.

\appendix

\section{Derivation of confidence intervals}
\label{ConfInt}

Starting from Equation~\eqref{condE}, the conditional expectation of $\hat R(t)$ given the past is
\eqn{ \E(\hat R(t) \cond I(t-1), \dots) = \frac{\mathbb E(I(t) \vert I(t-1), \dots)}{\sum_{\tau=1}^\infty w(\tau) I(t-\tau)} = R(t)\,. }
Therefore, $\hat R(t)$ is \term{unbiased},
\eqn{ \E \hat R(t) = R(t)\,, }
and the \term{conditional variance} of $\hat R(t)$ is given by
\eqn{ \Var(\hat R(t) \cond I(t-1), \dots) = \frac{R(t)}{\sum_{\tau=1}^\infty w(\tau) I(t-\tau)}\,.}
An application of Slutsky's lemma gives an asymptotic $(1-\alpha)$-confidence interval for $R(t)$: if $q$ denotes the $(1-\frac\alpha 2)$-quantile of the standard normal distribution it is given by
\eqn{ \Biggl[\hat R(t) - q \sqrt{\frac{\hat R(t)}{\sum_{\tau=1}^\infty w(\tau) I(t-\tau)}}, \hat R(t) + q \sqrt{\frac{\hat R(t)}{\sum_{\tau=1}^\infty w(\tau) I(t-\tau)}} \Biggr]\,. }
Note that (approximate) coverage is always guaranteed conditionally on the past, and hence also without conditioning.

\section{Derivation of the infectivity profile for the SEIR-model}
\label{InfProfSEIR}

Both latent period and infectious period are geometrically distributed with parameters $p_I$ and $p_R$, respectively.
We essentially need to compute the convolution (summing over time $s$ of getting infectious).
For $\tau > 1$ and assuming $p_I > p_R$ (the other cases are similar), we obtain
\begin{align}
w(\tau) &= \sum_{s=1}^{\tau - 1} p_I (1 - p_I)^{s-1} (1-p_R)^{\tau-1-s}
= p_I (1-p_R)^{\tau-2} \sum_{s=0}^{\tau - 2} \biggl( \frac{1 - p_I}{1-p_R} \biggr)^s
\notag\\&= p_I (1-p_R)^{\tau-2} \frac{1 - \bigl( \frac{1 - p_I}{1-p_R} \bigr)^{\tau-1}}{ \frac{p_I - p_R}{1 - p_R} }
= p_I (1-p_R)^{\tau-1} \frac{1 - \bigl( \frac{1 - p_I}{1-p_R} \bigr)^{\tau-1}}{ p_I - p_R }
\notag\\&= p_I \frac{(1-p_R)^{\tau-1} - (1 - p_I)^{\tau-1}}{ p_I - p_R }\,.
\end{align}

\bibliographystyle{dcu}
\setlength{\bibsep}{3pt}
\bibliography{../corona.bib}

\end{document}